\documentclass[twocolumn,aps,superscriptaddress,nofootinbib,floatfix,linenumbers]{revtex4}
\usepackage{amsfonts}
\usepackage{float}

\usepackage{placeins}
\usepackage{graphicx}
\usepackage[hidelinks]{hyperref}
\usepackage{color,amsmath,amssymb,bm}
\usepackage{tensor}
\usepackage[cmyk]{xcolor}

\usepackage{amssymb}
\usepackage{amsmath}

\begin{document}



\title{Heavy Quark Recombination Across Collision Systems: Meson and Baryon production}

\author{Vincenzo Minissale}
\affiliation{Department of Physics and Astronomy "E. Majorana", University of Catania, Via S. Sofia 64, 1-95123 Catania, Italy}
\affiliation{INFN Sezione di Catania, Via Santa Sofia,64 - 95123 Catania, Italy}

\author{Salvatore Plumari}
\affiliation{Department of Physics and Astronomy "E. Majorana", University of Catania, Via S. Sofia 64, 1-95123 Catania, Italy}
\affiliation{Laboratori Nazionali del Sud, INFN-LNS, Via S. Sofia 62, I-95123 Catania, Italy}

\author{Maria Lucia Sambataro}
\affiliation{Department of Physics and Astronomy "E. Majorana", University of Catania, Via S. Sofia 64, 1-95123 Catania, Italy}
\affiliation{Laboratori Nazionali del Sud, INFN-LNS, Via S. Sofia 62, I-95123 Catania, Italy}

\author{Gloria Calà}
\affiliation{Department of Physics and Astronomy "E. Majorana", University of Catania, Via S. Sofia 64, 1-95123 Catania, Italy}
\affiliation{Laboratori Nazionali del Sud, INFN-LNS, Via S. Sofia 62, I-95123 Catania, Italy}

\author{Vincenzo Greco}
\affiliation{Department of Physics and Astronomy "E. Majorana", University of Catania, Via S. Sofia 64, 1-95123 Catania, Italy}
\affiliation{Laboratori Nazionali del Sud, INFN-LNS, Via S. Sofia 62, I-95123 Catania, Italy}

%% Abstract
\begin{abstract}
Measurements of heavy-flavor baryons in nucleus--nucleus and proton--proton
collisions have provided strong evidence that heavy-quark hadronization cannot
be described solely in terms of vacuum fragmentation. In particular, the large
enhancement of the $\Lambda_c/D^0$, $\Xi_c/D^0$ and $\Omega_c/D^0$ ratios observed
at RHIC and LHC energies has been successfully described within a coalescence
plus fragmentation approach, supporting the relevance of recombination mechanisms
in both large and small collision systems.
Building on these results, we discuss recent developments in the coalescence
hadronization framework. Particular emphasis is given to the production of
multi-charmed baryons, namely $\Xi_{cc}$, and $\Omega_{ccc}$,
whose yields are predicted in a wide range of collision systems from
PbPb to KrKr, ArAr and OO collisions. The strong sensitivity of these states,
especially $\Omega_{ccc}$, to the underlying charm-quark distribution makes them
valuable probes of charm thermalization and non-equilibrium effects in the QGP.
We also present recent extensions of the model to the beauty sector,
including predictions for $\Lambda_b/B^0$, $\Xi_b/B^0$ and $\Omega_b/B^0$
production. The results indicate sizeable recombination effects and a stronger
baryon enhancement than in the charm sector, highlighting the increasing role
of coalescence for heavier quarks.
Finally, we briefly comment on ongoing developments toward a unified description
of open and hidden heavy flavor through the application of the same framework to
quarkonium production in small collision systems.

\end{abstract}
\maketitle

\section{Introduction}

Heavy quarks constitute one of the most powerful probes of strongly interacting matter created in high-energy nuclear collisions. Owing to their large masses, charm and bottom quarks are predominantly produced during the initial hard scatterings and therefore experience the entire evolution of the produced medium. Their final-state distributions carry information on both the transport properties of the Quark-Gluon Plasma (QGP) and the mechanisms governing hadron formation.
Experimental measurements performed at RHIC and LHC have demonstrated a strong coupling of heavy quarks with the medium. The observed suppression of open heavy-flavor hadrons at high transverse momentum and the sizable anisotropic flow measured in heavy-ion collisions indicate that heavy quarks participate in the collective expansion of the QGP. These observations have motivated extensive theoretical efforts to understand the interplay between transport and hadronization dynamics \cite{
Gossiaux:2008jv,He:2012df,Cao:2015hia,Sambataro:2022sns,Sambataro:2023tlv,Sambataro:2025obe,Plumari:2015cfa, Scardina:2017ipo, Plumari:2019hzp, Sun:2019gxg,Oliva:2020doe}.
One of the main open questions concerns the hadronization process of heavy quarks. In elementary collisions, heavy-flavor production is commonly described within perturbative QCD supplemented by fragmentation functions extracted from $e^{+}e^{-}$ experiments. However, recent measurements of heavy-flavor baryon-to-meson ratios have challenged this paradigm. In particular, the $\Lambda_c/D^0$, $\Xi_c/D^0$ and $\Omega_c/D^0$ ratios measured at the LHC significantly exceed expectations based on fragmentation alone, suggesting that additional hadronization mechanisms contribute to particle production.
A natural explanation is provided by quark recombination, or coalescence, where nearby quarks in phase space combine into hadronic states during the hadronization stage. Coalescence-based approaches have successfully explained several features of the light-flavor sector and have become increasingly important for the interpretation of heavy-flavor measurements \cite{Minissale:2020bif, Beraudo:2023nlq, Zhao:2023ucp,Zhao:2023nrz}.
We have developed a hybrid hadronization framework that combines coalescence and fragmentation mechanisms. The model has successfully described heavy-flavor observables in nucleus--nucleus collisions and has recently been extended to investigate multicharm hadrons, bottom baryons in proton--proton collisions, and quarkonium formation in small systems. 

%%%%%%%%%%%%%%%%%%%%%%%%%%%%%%%%%%%%%%%%%%%%%%%%%%%%%%%%%%%%%%

\section{Coalescence plus Fragmentation Framework}

Our approach describes hadronization through the interplay between quark recombination and fragmentation. During hadronization, heavy quarks may coalesce with nearby light quarks according to Wigner distributions derived from hadronic wave functions. Heavy quarks that do not satisfy the coalescence criterion subsequently hadronize through standard fragmentation functions.
The coalescence production is obtained from
\begin{eqnarray}
\label{eq-coal}
\frac{dN_{H}}{dyd^{2}P_{T}} = g_{H} \int \prod^{N_{q}}_{i=1} \frac{d^{3}p_{i}}{(2\pi)^{3}E_{i}} p_{i} \cdot d\sigma_{i}  \; f_{q_i}(x_{i}, p_{i}) \times \nonumber \\ 
 C_{H}(x_{1}...x_{N_{q}}, p_{1}...p_{N_{q}})\, \delta^{(3)} \left(P-\sum^{n}_{i=1} p_{i} \right) \nonumber
\end{eqnarray}

The function $C_{H}(x_{1}...x_{N_{q}})=C^{N_q-1}f_{H}(x_{1}...x_{N_{q}}, p_{1}...p_{N_{q}})$ is expressed in terms of the Wigner function $f_{H}(x_{1}...x_{N_{q}}, p_{1}...p_{N_{q}})$; where $C^{N_q-1}$ is a normalization factor determined by imposing the coalescence probability equal to one when the quark momentum is zero. The widths $\sigma_{ri}$  of the Gaussian Wigner function are connected to the the root mean square charge radius of the hadron produced, for details see refs. \cite{Plumari:2017ntm,Minissale:2023dct,Minissale:2020bif,Minissale:2024gxx}. Ground states and first excited resonances are included together with their feed-down contributions.
In heavy-ion collisions, charm-quark distributions are obtained from relativistic transport calculations, while light quarks are described through thermal distributions incorporating collective radial flow. This framework allows a unified description of hadron production over a wide range of collision energies and system sizes \cite{Minissale:2015zwa}.

%%%%%%%%%%%%%%%%%%%%%%%%%%%%%%%%%%%%%%%%%%%%%%%%%%%%%%%%%%%%%%

\section{Results}

The production of hadrons containing multiple charm quarks can be used to probe the charm-quark degree of thermalization in the QGP. More than conventional open heavy-flavor hadrons, multicharm states strongly depend on the local charm density in phase space at hadronization and therefore provide valuable information on heavy-quark transport.
In our analysis we have investigated the production for several collision systems, including Pb-Pb, Kr-Kr, Ar-Ar and O-O collisions, allowing a systematic investigation of system-size effects.
Particular attention was devoted to the triply charmed baryon $\Omega_{ccc}$. Since its formation requires the recombination of three charm quarks, the production yield is extremely sensitive to the charm distributions at hadronization. The calculations show that realistic non-equilibrium charm distributions lead to a sizeable suppression of the $\Omega_{ccc}$ yield with respect to thermal-equilibrium scenarios.
A further advantage of the coalescence approach is the possibility of predicting transverse-momentum spectra. The resulting momentum distributions display characteristic differences with respect to statistical-hadronization calculations, opening the possibility of experimentally discriminating among different hadronization mechanisms \cite{Minissale:2023dct,Andronic:2021erx}.
%%%%%%%%%%%%%%%%%%%%%%%%%%%%%%%%%%%%%%%%%%%%%%%%%%%%%%%%%%%%%%
Recent measurements of heavy-flavor baryons in proton--proton collisions have revealed unexpectedly large baryon-to-meson ratios. These observations challenge the traditional picture based exclusively on fragmentation and suggest that recombination effects may contribute to hadron formation even in small systems.
Motivated by these experimental findings, the coalescence plus fragmentation framework has recently been extended to the beauty sector \cite{Oh:2009zj, Plumari:2017ntm,Minissale:2020bif}.
One of the most relevant results is that bottom-quark coalescence remains sizable over a much broader transverse-momentum region than in the charm sector. When expressed in terms of equal heavy-quark velocity, charm and bottom exhibit remarkably similar recombination probabilities.
The calculations predict that bottom-baryon production receives a dominant contribution from coalescence up to transverse momenta of approximately 10 GeV/$c$. Consequently, substantial enhancements of baryon-to-meson ratios are obtained relative to fragmentation-only expectations. In particular, the predicted $\Lambda_b/B^0$ ratio is significantly enhanced and is found to be approximately 50\% larger than the corresponding $\Lambda_c/D^0$ ratio in the low and intermediate-momentum region. Similar behaviors are obtained for the $\Xi_b/B^0$ ratio, while predictions for the $\Omega_b/B^0$ production provide an interesting results for future experimental measurements comparison.

\begin{figure}[t]
\centering
\includegraphics[width=0.5\linewidth]{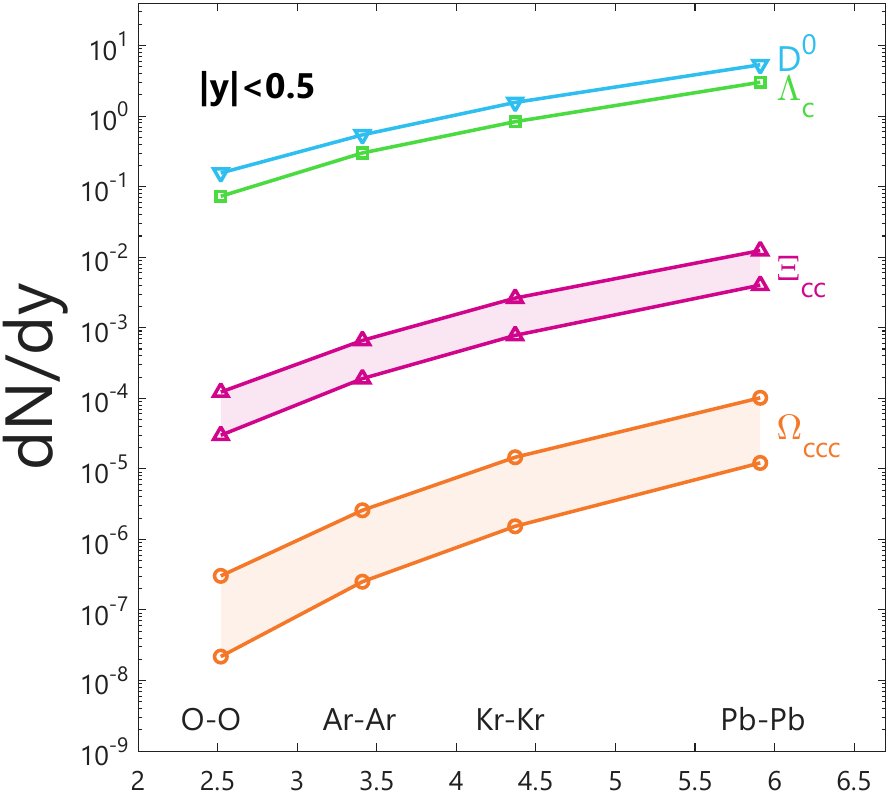}
  \caption{Mid-rapidity $dN/dy$ for single- and multi-charmed hadrons in various collision systems, calculated using the coalescence-plus-fragmentation approach. Error bands indicate the yield range, where the lower bound corresponds to a realistic charm-quark distribution and the upper bound assumes full thermalization. }
\label{Fig1}
\end{figure}

\begin{figure}[t]
\centering
\includegraphics[width=0.75\linewidth]{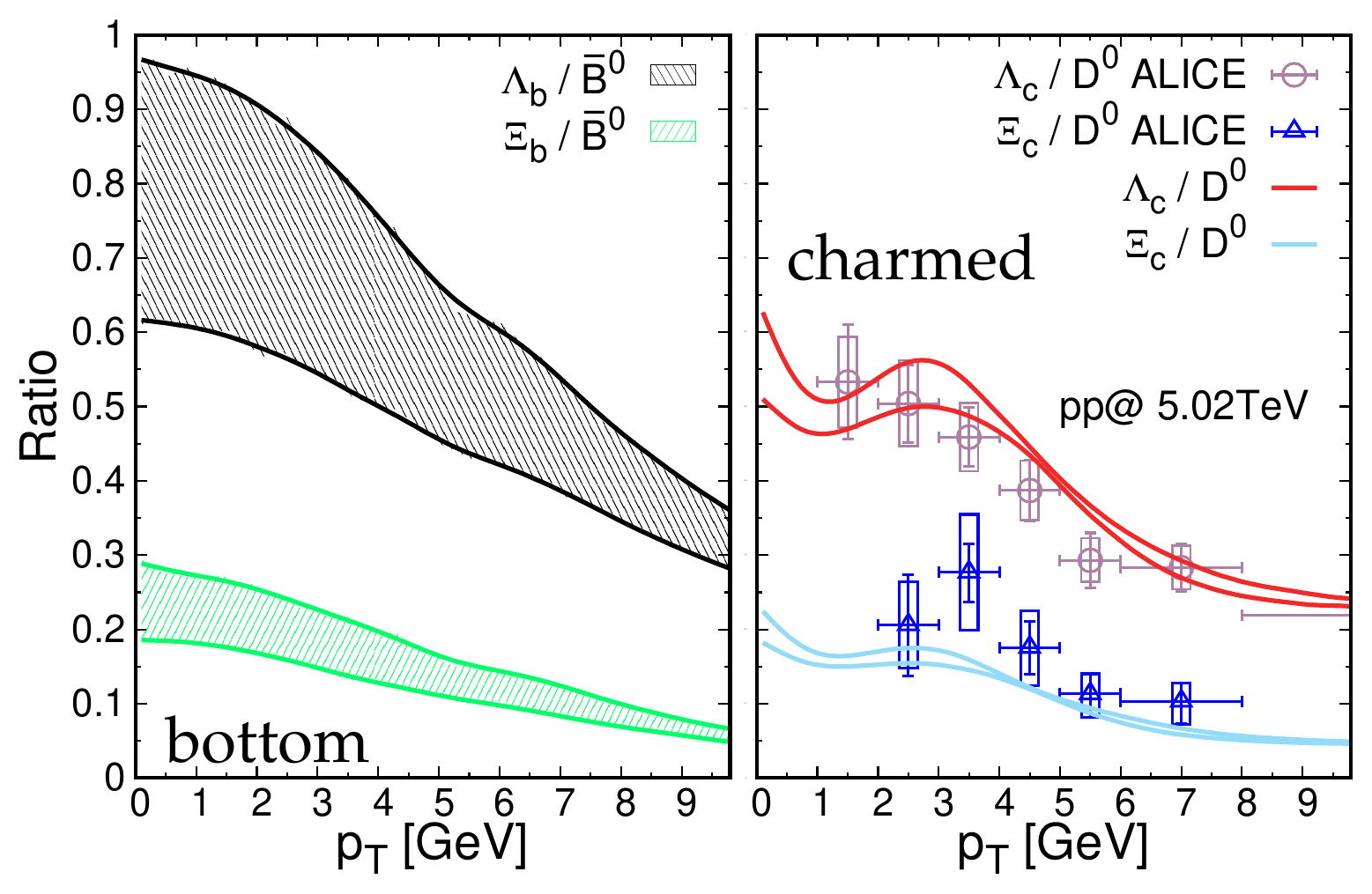}
\caption{Baryon-to-meson ratios as a function of transverse momentum for $pp$ collisions at $\sqrt{s}=5.02~\text{TeV}$: (left) $\Lambda_c/D^0$, $\Xi_c/D^0$ and (right) $\Lambda_b/B^0$, $\Xi_b/B^0$. The shaded bands represent variations in the Wigner width. Experimental data are taken from~\cite{ALICE:2020wla, ALICE:2021psx}.
}
\label{Fig2}
\end{figure}
%%%%%%%%%%%%%%%%%%%%%%%%%%%%%%%%%%%%%%%%%%%%%%%%%%%%%%%%%%%%%%

A further recent development concerns the extension of the recombination framework to hidden heavy flavor. In this approach, heavy quarkonia are formed through the coalescence of heavy quark--antiquark pairs described by Wigner functions constructed from solutions of the Schr\"odinger equation.
Both harmonic-oscillator and Cornell potentials have been investigated for the description of quarkonium wave functions. 
Two different scenarios have been considered. The first assumes direct hadronization of heavy quarks distributed according to FONLL spectra. The second incorporates an in-medium evolution inside a short-lived QGP fireball before recombination occurs.
The inclusion of medium effects leads to an improved description of the low-transverse-momentum spectra of both $J/\psi$ and $\Upsilon(1S)$ production. These results suggest that collective phenomena may influence heavy-quark dynamics even in proton--proton collisions.

%%%%%%%%%%%%%%%%%%%%%%%%%%%%%%%%%%%%%%%%%%%%%%%%%%%%%%%%%%%%%%

%%%%%%%%%%%%%%%%%%%%%%%%%%%%%%%%%%%%%%%%%%%%%%%%%%%%%%%%%%%%%%


\begin{thebibliography}{33}

\bibitem{Gossiaux:2008jv}
P.B. Gossiaux, J.~Aichelin, 
%{Towards an understanding of the RHIC single electron data}, 
Phys. Rev. C \textbf{78}, 014904 (2008).%, \texttt{0802.2525}. %%%\doiwoc{10.1103/PhysRevC.78.014904}

\bibitem{He:2012df}
M.~He, R.J. Fries, R.~Rapp, %{$\mathbf{D_s}$-Meson as Quantitative Probe of Diffusion and Hadronization in Nuclear Collisions}, 
Phys. Rev. Lett. \textbf{110}, 112301 (2013).%, \texttt{1204.4442}. %%\doiwoc{10.1103/PhysRevLett.110.112301}


\bibitem{Cao:2015hia}
S.~Cao, G.Y. Qin, S.A. Bass, %{Energy loss, hadronization and hadronic interactions of heavy flavors in relativistic heavy-ion collisions}, 
Phys. Rev. \textbf{C92}, 024907 (2015).%, \texttt{1505.01413}. %%\doiwoc{10.1103/PhysRevC.92.024907}


\bibitem{Sambataro:2022sns}
M.L. Sambataro, Y.~Sun, V.~Minissale, S.~Plumari, V.~Greco, %{Event-shape engineering analysis of D meson in ultrarelativistic heavy-ion collisions}, 
Eur. Phys. J. C \textbf{82}, 833 (2022).%, \texttt{2206.03160}. %%\doiwoc{10.1140/epjc/s10052-022-10802-2}

\bibitem{Sambataro:2023tlv}
M.L. Sambataro, V.~Minissale, S.~Plumari, V.~Greco, 
%{B meson production in Pb+Pb at 5.02 ATeV at LHC: Estimating the diffusion coefficient in the infinite mass limit}, 
Phys. Lett. B \textbf{849}, 138480 (2024).%, \texttt{2304.02953}. %%\doiwoc{10.1016/j.physletb.2024.138480}

%\cite{Sambataro:2025obe}
\bibitem{Sambataro:2025obe}
M.~L.~Sambataro, V.~Minissale, S.~Plumari and V.~Greco,
%``Assessing the lattice QCD space diffusion coefficient and the thermalization time of charm quark by mean of D meson observables at LHC,''
arXiv:2508.01024.

\bibitem{Plumari:2015cfa}
S.~Plumari, G.L. Guardo, F.~Scardina, V.~Greco, 
%{Initial state fluctuations from mid-peripheral to ultra-central collisions in a event-by-event transport approach}, 
Phys. Rev. \textbf{C92}, 054902 (2015).%, \texttt{1507.05540}. %%\doiwoc{10.1103/PhysRevC.92.054902}

\bibitem{Scardina:2017ipo}
F.~Scardina, S.K. Das, V.~Minissale, S.~Plumari, V.~Greco, 
%{Estimating the charm quark diffusion coefficient and thermalization time from D meson spectra at energies available at the BNL Relativistic Heavy Ion Collider and the CERN Large Hadron Collider}, 
Phys. Rev. \textbf{C96}, 044905 (2017).%, \texttt{1707.05452}. %%\doiwoc{10.1103/PhysRevC.96.044905}

\bibitem{Plumari:2019hzp}
S.~Plumari, G.~Coci, V.~Minissale, S.K. Das, Y.~Sun, V.~Greco, 
%{Heavy - light flavor correlations of anisotropic flows at LHC energies within event-by-event transport approach}, 
Phys. Lett. B \textbf{805}, 135460 (2020).%, \texttt{1912.09350}. %%\doiwoc{10.1016/j.physletb.2020.135460}

\bibitem{Sun:2019gxg}
Y.~Sun, S.~Plumari, V.~Greco, 
%{Study of collective anisotropies v2 and v3 and their fluctuations in pA collisions at LHC within a relativistic transport approach}, 
Eur. Phys. J. C \textbf{80}, 16 (2020).%, \texttt{1907.11287}. %\doiwoc{10.1140/epjc/s10052-019-7577-7}

\bibitem{Oliva:2020doe}
L.~Oliva, S.~Plumari, V.~Greco, %{Directed flow of D mesons at RHIC and LHC: non-perturbative dynamics, longitudinal bulk matter asymmetry and electromagnetic fields}, 
JHEP \textbf{05}, 034 (2021).%, \texttt{2009.11066}. %\doiwoc{10.1007/JHEP05(2021)034}


\bibitem{Minissale:2020bif}
V.~Minissale, S.~Plumari, V.~Greco, 
%{Charm hadrons in pp collisions at LHC energy within a coalescence plus fragmentation approach}, 
Phys. Lett. B \textbf{821}, 136622 (2021).%, \texttt{2012.12001}. %\doiwoc{10.1016/j.physletb.2021.136622}

\bibitem{Beraudo:2023nlq}
A.~Beraudo, A.~De~Pace, D.~Pablos, F.~Prino, M.~Monteno, M.~Nardi, %{Heavy-flavor transport and hadronization in pp collisions}, 
Phys. Rev. D \textbf{109}, L011501 (2024).%, \texttt{2306.02152}. %\doiwoc{10.1103/PhysRevD.109.L011501}

\bibitem{Zhao:2023ucp}
J.~Zhao, J.~Aichelin, P.B. Gossiaux, K.~Werner, %{Heavy flavor as a probe of hot QCD matter produced in proton-proton collisions}, 
Phys. Rev. D \textbf{109}, 054011 (2024).%, \texttt{2310.08684}. %\doiwoc{10.1103/PhysRevD.109.054011}


\bibitem{Zhao:2023nrz}
J.~Zhao, J.~Aichelin, P.~B.~Gossiaux, A.~Beraudo, S.~Cao, W.~Fan, M.~He, V.~Minissale, T.~Song and I.~Vitev, \textit{et al.}
%``Hadronization of heavy quarks,''
Phys. Rev. C \textbf{109} (2024) no.5, 054912


\bibitem{Plumari:2017ntm}
S.~Plumari, V.~Minissale, S.K. Das, G.~Coci, V.~Greco, %{Charmed Hadrons from Coalescence plus Fragmentation in relativistic nucleus-nucleus collisions at RHIC and LHC}, 
Eur. Phys. J. \textbf{C78}, 348 (2018).%, \texttt{1712.00730}. %\doiwoc{10.1140/epjc/s10052-018-5828-7}


\bibitem{Minissale:2023dct}
V.~Minissale, S.~Plumari, Y.~Sun, V.~Greco, %{Multi-charmed and singled charmed hadrons from coalescence: yields and ratios in different collision systems at LHC}, 
Eur. Phys. J. C \textbf{84}, 228 (2024), \texttt{2305.03687}. %\doiwoc{10.1140/epjc/s10052-024-12571-6}

\bibitem{Minissale:2024gxx}
V.~Minissale, V.~Greco and S.~Plumari,
%``Bottomed mesons and baryons production in pp collisions at s=5 TeV LHC energy within a Coalescence plus Fragmentation approach,''
Phys. Lett. B \textbf{860} (2025), 139190

\bibitem{Minissale:2015zwa}
V.~Minissale, F.~Scardina, V.~Greco, %{Hadrons from coalescence plus fragmentation in AA collisions at energies available at the BNL Relativistic Heavy Ion Collider to the CERN Large Hadron Collider}, 
Phys. Rev. C \textbf{92}, 054904 (2015), \texttt{1502.06213}. %\doiwoc{10.1103/PhysRevC.92.054904}

\bibitem{Oh:2009zj}
Y.~Oh, C.M. Ko, S.H. Lee, S.~Yasui, %{Heavy baryon/meson ratios in relativistic heavy ion collisions}, 
Phys. Rev. C \textbf{79}, 044905 (2009), \texttt{0901.1382}. %\doiwoc{10.1103/PhysRevC.79.044905}

\bibitem{Andronic:2021erx}
A.~Andronic, et~al.
%P.~Braun-Munzinger, M.K. K\"ohler, A.~Mazeliauskas, K.~Redlich, J.~Stachel, V.~Vislavicius, %{The multiple-charm hierarchy in the statistical hadronization model}, 
JHEP \textbf{07}, 035 (2021).%, \texttt{2104.12754}. %\doiwoc{10.1007/JHEP07(2021)035}


\bibitem{ALICE:2020wla}
S.~Acharya et~al. (ALICE), %{$\Lambda^+_c$ production in $pp$ and in $p$-Pb collisions at $\sqrt {s_{NN}}$=5.02 TeV}, 
Phys. Rev. C \textbf{104}, 054905 (2021), \texttt{2011.06079}. %\doiwoc{10.1103/PhysRevC.104.054905}

\bibitem{ALICE:2021psx}
S.~Acharya \textit{et al.} [ALICE],
%``Measurement of the production cross section of prompt $ {\Xi}_{\mathrm{c}}^0 $ baryons at midrapidity in pp collisions at $ \sqrt{s} $ = 5.02 TeV,''
JHEP \textbf{10} (2021), 159
doi:10.1007/JHEP10(2021)159

\end{thebibliography}
\end{document}